\documentclass{article}
\usepackage{amssymb}
\usepackage{amsmath}

\begin{document}

\begin{center}
\textbf{Interesting system of }$3$ first-order recursions

\bigskip 

\textbf{Francesco Calogero}

Physics Department, University of Rome "La Sapienza", Rome, Italy

Istituto Nazionale di Fisica Matematica, Sezione di Roma 1, Rome, Italy

Istituto Nazionale di Alta Matematica, Gruppo Nazionale di Fisica
Matematica, Italy

francesco.calogero@uniroma1.it, francesco.calogero@roma1.infn.it

\bigskip

\textbf{Abstract}

\bigskip
\end{center}

In this paper we firstly review how to \textit{explicitly} solve a system of 
$3$ \textit{first-order linear recursions }and outline the main properties
of these solutions. Next, via a change of variables, we identify a class of
systems of $3$ \textit{first-order nonlinear recursions} which also are 
\textit{explicitly solvable}. These systems might be of interest for
practitioners in \textit{applied} sciences: they allow a complete display of
their solutions, which may feature interesting behaviors, for instance be 
\textit{completely periodic} ("isochronous systems", if the independent
variable $n=0,1,2,3...$is considered a \textit{ticking time}), or feature
this property \textit{only asymptotically} (as\textit{\ }$n\rightarrow
\infty $).

\bigskip

\textbf{Introduction}

\bigskip

In this paper we investigate systems of $3$ \textit{first-order} recursions.
We start from \textit{linear} systems. Such systems of recursions are 
\textit{explicitly} solvable via \textit{algebraic} operations, but the
results are generally rather complicated, hence it makes sense to try and
identify interesting solutions which are somewhat manageable and display
interesting evolutions, in the \textit{discrete independent} variable $n$
taking the \textit{integer} values $n=0,1,2,...$(a \textit{ticking time}).
We then use these findings to identify certain classes of $3$ \textit{%
first-order} \textit{nonlinear }recursions which are as well \textit{%
explicitly} solvable via \textit{algebraic} operations hence also perhaps
interesting (for instance, in \textit{applicative} contexts).

\bigskip

\textbf{The system of }$3$\textbf{\ linear recurrences}

\bigskip

The \textit{general autonomous linear} system of $3$ \textit{first-order }%
recursions reads as follows:%
\begin{equation}
z_{j}\left( n+1\right) =\overline{z}_{j}+\sum_{k=1}^{3}\left[
a_{jk}z_{k}\left( n\right) \right] ~,~~\ j=1,2,3,~~\ n=0,1,2,...~.
\label{3Recsz}
\end{equation}

\textbf{Notation}. Hereafter the indices $j$ and $k$ take the $3$ values $%
1,2,3,$ and the index $\ell $ (used later) take the $4$ values $\ell
=0,1,2,3 $; the $9$ coefficients $a_{jk}$ are \textit{real} numbers (\textit{%
a priori} arbitrarily assigned, but possibly subjected to some restrictions,
see below); and we also assume the $3$ \textit{dependent} variables $%
z_{j}\left( n\right) $ to be \textit{real} numbers, as well as the $3$
constants $\overline{z}_{j}$; but of course in the following computations we
may need to also use \textit{complex} numbers, and in that context we shall
use the notation $\mathbf{i}$ to denote the \textit{imaginary unit, }hence $%
\mathbf{i}^{2}=-1$. $\blacksquare $

The first standard---quite trivial---development is to reduce the system (%
\ref{3Recsz}) to the simpler \textit{homogeneous} version%
\begin{equation}
y_{j}\left( n+1\right) =\sum_{k=1}^{3}\left[ a_{jk}y_{k}\left( n\right) %
\right] ~,~~\ j=1,2,3,~~\ n=0,1,2,...~.  \label{3Recs}
\end{equation}%
To perform this simple step it is convenient to introduce the vector-matrix
notation, by setting 
\begin{subequations}
\begin{eqnarray}
\mathbf{z}\left( n\right) &\equiv &\left( z_{1}\left( n\right)
\,,z_{2}\left( n\right) \,,z_{3}\left( n\right) \,\right) ~, \\
\overline{\mathbf{z}} &=&\left( \overline{z}_{1}\,,\overline{z}_{2}\,,%
\overline{z}_{3}\right) ~, \\
\mathbf{y}\left( n\right) &=&\left( y_{1}\left( n\right) \,,y_{2}\left(
n\right) \,,y_{3}\left( n\right) \,\right) ~, \\
\overline{\mathbf{y}} &\mathbf{=}&\left( \overline{y}_{1},\overline{y}_{2},%
\overline{y}_{3}\right) ~, \\
\mathbf{z}\left( n\right) &=&\overline{\mathbf{y}}+\mathbf{y}\left( n\right)
~, \\
\mathbf{A} &\equiv &\left( 
\begin{array}{ccc}
a_{11} & a_{12} & a_{13} \\ 
a_{21} & a_{22} & a_{23} \\ 
a_{31} & a_{32} & a_{33}%
\end{array}%
\right) ~.
\end{eqnarray}%
Then the $3$ recursions (\ref{3Recs}) read as the following single $3$%
-vector recursion 
\end{subequations}
\begin{subequations}
\begin{equation}
\mathbf{z}\left( n+1\right) =\overline{\mathbf{z}}+\mathbf{Az}\left(
n\right) ~,
\end{equation}%
and by setting%
\begin{equation}
\overline{\mathbf{y}}=\overline{\mathbf{z}}+\mathbf{A}\overline{\mathbf{y}}~,
\end{equation}%
implying%
\begin{equation}
\overline{\mathbf{y}}=\left( 1-\mathbf{A}\right) ^{-1}\overline{z}~,
\end{equation}%
we get for the $3$-vector $\mathbf{y}\left( n\right) $ the \textit{%
homogeneous linear} recursion (\ref{3Recs}) (in its $3$-vector version).

So we hereafter focus on the system (\ref{3Recs}) of $3$ \textit{linear
homogeneous} recursions.

\bigskip

\textbf{A special solution}

\bigskip

Let us set 
\end{subequations}
\begin{subequations}
\label{SpecSol}
\begin{equation}
y_{j}\left( n\right) =\gamma _{j}u^{n}~,
\end{equation}%
where $\gamma _{j}$ and $u$ are $3+1=4$ \textit{a priori arbitrary} numbers.
Then eq. (\ref{3Recs}) implies 
\begin{equation}
\gamma _{j}u-\sum_{k=1}^{3}\left( a_{jk}\gamma _{k}\right) =0~,~~j=1,2,3~.
\end{equation}%
This system of $3$ \textit{linear algebraic} equations for the $3$ numbers $%
\gamma _{j}$ would imply that they \textit{all vanish}, unless the \textit{%
determinant} of the coefficients of the $3$ numbers $\gamma _{j}$ \textit{%
vanishes;} so we must enforce the condition 
\end{subequations}
\begin{equation}
\left\vert 
\begin{array}{ccc}
u-a_{11} & -a_{12} & -a_{13} \\ 
-a_{21} & u-a_{22} & -a_{23} \\ 
-a_{31} & -a_{32} & u-a_{33}%
\end{array}%
\right\vert =0~,
\end{equation}%
which amounts to the following \textit{cubic} equation for the number $u$: 
\begin{subequations}
\label{uAAA}
\begin{equation}
u^{3}+A_{2}u^{2}+A_{1}u+A_{0}=0~,  \label{CubicEqu}
\end{equation}%
where 
\begin{eqnarray}
A_{2} &=&-\left( a_{11}+a_{22}+a_{33}\right) ~,  \label{A2} \\
A_{1}
&=&a_{11}a_{22}+a_{22}a_{33}+a_{33}a_{11}-a_{12}a_{21}-a_{13}a_{31}-a_{23}a_{32}~,
\label{A1} \\
A_{0} &=&-a_{11}a_{22}a_{33}-a_{12}a_{23}a_{31}-a_{13}a_{21}a_{32}+  \notag
\\
&&a_{13}~a_{31}a_{22}+a_{12}a_{21}a_{33}+a_{23}a_{32}a_{11}~.  \label{A0}
\end{eqnarray}%
So hereafter we assume that $u$ is one of the $3$ roots of the cubic eq. (%
\ref{CubicEqu}). Then one of the $3$ parameters $\gamma _{j}$ may be \textit{%
arbitrarily} assigned (say, $\gamma _{1}$), while the other $2$ are
determined as the solution of the following system of $2$ \textit{algebraic}
equations: 
\end{subequations}
\begin{subequations}
\label{Eqsgam}
\begin{eqnarray}
\left( u-a_{22}\right) \gamma _{2}-a_{23}\gamma _{3} &=&a_{21}\gamma _{1}~,
\label{Eq1gam1} \\
-a_{32}\gamma _{2}+\left( u-a_{33}\right) \gamma _{3} &=&a_{31}\gamma _{1}~,
\label{Eq2gam1}
\end{eqnarray}%
implying%
\begin{eqnarray}
\gamma _{2} &=&\frac{\gamma _{1}\left[ \left( u-a_{33}\right)
a_{21}+a_{23}a_{31}\right] }{\left( u-a_{22}\right) \left( u-a_{33}\right)
-a_{23}a_{32}}~,  \label{gam20} \\
\gamma _{3} &=&\frac{\gamma _{1}\left[ \left( u-a_{22}\right)
a_{31}+a_{32}a_{21}\right] }{\left( u-a_{22}\right) \left( u-a_{33}\right)
-a_{23}a_{31}}~.  \label{gam30}
\end{eqnarray}%
So, the $2$ numbers $\gamma _{2}$ and $\gamma _{3}$ are proportional to the
number $\gamma _{1}$, with the proportionality factor depending on the value
of the root $u$.

As for the $3$ roots $u_{0}$, $u_{+}$, $u_{-}$ of the \textit{cubic}
equation (\ref{CubicEqu}), we may explicitly write them---with some help
from \textbf{Mathematica} (which actually essentially uses the so-called
Cardano formulas)---as follows: 
\end{subequations}
\begin{subequations}
\label{SoluzHomo}
\begin{equation}
u_{0}=\left[ -A_{2}+2^{1/3}\left( A_{2}^{2}-2A_{1}\right) /E+2^{-1/3}E\right]
/3~,  \label{u00}
\end{equation}%
\begin{equation}
u_{\pm }=G\pm \mathbf{i}\left( -1\right) ^{\ell }H_{-}~,~~~\ell =1,2~,
\label{uel}
\end{equation}%
\begin{equation}
G=-A_{2}+H_{+}~,  \label{G}
\end{equation}%
\begin{equation}
H_{\pm }=-\sqrt{3}\left[ \pm 2^{-4/3}E+2^{-2/3}\left(
A_{2}^{2}-3A_{1}\right) /E\right] ~,  \label{Hpm}
\end{equation}%
\begin{equation}
E=\left( -2A_{2}^{3}+9A_{2}A_{1}-27A_{0}+3F\right) ^{1/3}~,  \label{E}
\end{equation}%
\begin{equation}
F=\sqrt{3\left[ -\left( A_{2}A_{1}\right) ^{2}+4\left(
A_{1}^{3}+A_{2}^{3}A_{0}\right) -18A_{2}A_{1}A_{0}+27A_{0}^{2}\right] }~.
\label{F}
\end{equation}%
These formulas seem explicit, but are quite complicated, and in fact they
are somewhat undetermined since the values of \textit{quadratic} and \textit{%
cubic} roots are \textit{not} uniquely determined.

So we now try and introduce an alternative notation more convenient for our
purposes. Let 
\end{subequations}
\begin{subequations}
\begin{eqnarray}
&&0=u^{3}+A_{2}u^{2}+A_{1}u+A_{0}= \\
&&\left( u-\overline{u}\right) \left[ u-\widetilde{u}\exp \left( \mathbf{i}%
\pi r\right) \right] \left[ u-\widetilde{u}\exp \left( -\mathbf{i}\pi
r\right) \right] = \\
&&\left( u-\overline{u}\right) \left[ u^{2}-2u\widetilde{u}\cos \left( \pi
r\right) +\left( \widetilde{u}\right) ^{2}\right] ~,
\end{eqnarray}%
implying 
\end{subequations}
\begin{equation}
A_{2}=-\overline{u}-2\widetilde{u}\cos \left( \pi r\right) ~,~~A_{1}=\left( 
\widetilde{u}\right) ^{2}+2\overline{u}\widetilde{u}\cos \left( \pi r\right)
~,~~A_{0}=-\overline{u}\left( \widetilde{u}\right) ^{2}~;  \label{A210}
\end{equation}%
we are therefore replacing the $3$ roots $u_{0}$, $u_{+}$, $u_{-}$ of the 
\textit{cubic} equation (\ref{CubicEqu}) with the $3$ parameters $\overline{u%
}$, $\widetilde{u}$ and $r$, related to them as follows: 
\begin{subequations}
\begin{equation}
u_{0}=\overline{u}~,~~~u_{\pm }=\widetilde{u}\exp \left( \pm \mathbf{i}\pi
r\right)
\end{equation}%
implying%
\begin{equation}
\overline{u}=u_{0}~,~~~\left( \widetilde{u}\right) ^{2}=u_{+}u_{-}~,~~~r=\pi
^{-1}\arccos \left( \frac{u_{+}+u_{-}}{2\widetilde{u}}\right) ~.
\end{equation}

Note that we are clearly focusing on the, more interesting, case in which
the $3$ roots of the \textit{cubic} equation (\ref{CubicEqu}) are \textit{not%
} \textit{all} $3$ \textit{real}, but rather one ("$u_{0}=\overline{u}$") is 
\textit{real} while the other $2$ ("$u_{\pm }=\widetilde{u}\exp \left( \pm 
\mathbf{i}\pi r\right) $") are \textit{complex conjugate} of each other; of
course here and hereafter $\widetilde{u}$ and $r$ are also $2$ \textit{real }%
numbers, with $r$ satisfying (without loss of generality) the restriction 
\begin{equation}
0<r<1~.  \label{r}
\end{equation}%
This entails the following $3$ relations among the $9$ \textit{coefficients} 
$a_{jk}$ of the system of recursions (\ref{3Recs}) and the $3$ \textit{real}
parameters $\overline{u},\widetilde{u}$ and $r$: \ 
\end{subequations}
\begin{subequations}
\label{ajkuur}
\begin{eqnarray}
&&a_{11}+a_{22}+a_{33}=\overline{u}+2\widetilde{u}\cos \left( \pi r\right) ~,
\\
&&a_{11}a_{22}+a_{22}a_{33}+a_{33}a_{11}-a_{12}a_{21}-a_{13}a_{31}-a_{23}a_{32}=
\notag \\
&&\left( \widetilde{u}\right) ^{2}+2\overline{u}\widetilde{u}\cos \left( \pi
r\right) ~, \\
&&a_{11}a_{22}a_{33}+a_{12}a_{23}a_{31}+a_{13}a_{21}a_{32}-  \notag \\
&&a_{13}~a_{31}a_{22}-a_{12}a_{21}a_{33}-a_{23}a_{32}a_{11}=\overline{u}%
\left( \widetilde{u}\right) ^{2}~.
\end{eqnarray}

\bigskip

\textbf{The general solution}

\bigskip

Then the \textit{general solution} of the system of $3$ \textit{linear}
recursions (\ref{3Recs}) reads as follows: 
\end{subequations}
\begin{subequations}
\label{yjngam}
\begin{equation}
y_{j}\left( n\right) =\gamma _{j}\left( \overline{u}\right) ^{n}+\gamma
_{j+}\left( \widetilde{u}\right) ^{n}\exp \left( \mathbf{i}n\pi r\right)
+\gamma _{j-}\left( \widetilde{u}\right) ^{n}\exp \left( -\mathbf{i}n\pi
r\right) ~,  \label{yngam}
\end{equation}%
or equivalently%
\begin{equation}
y_{j}\left( n\right) =\gamma _{j}\left( \overline{u}\right) ^{n}+\left(
\gamma _{j+}+\gamma _{j-}\right) \left( \widetilde{u}\right) ^{n}\cos \left(
nr\pi \right) +\mathbf{i}\left( \gamma _{j+}-\gamma _{j-}\right) \left( 
\widetilde{u}\right) ^{n}\sin \left( nr\pi \right) ~.
\end{equation}

Since we prefer to work with \textit{real} numbers, we now set (without loss
of generality) 
\end{subequations}
\begin{subequations}
\label{alfbetgam}
\begin{equation}
\gamma _{j+}+\gamma _{j-}=\alpha _{j}~,~~~\gamma _{j+}-\gamma _{j-}=-\mathbf{%
i}\beta _{j}~,  \label{alfabeta}
\end{equation}%
or, equivalently%
\begin{equation}
\alpha _{j}=2 \Re \left[ \gamma _{j+}\right] ~,~~~\beta _{j}=-2\Im%
\left[ \gamma _{j+}\right] ~;
\end{equation}%
where $ \Re $ and $\Im$ denote the real and imaginary parts of their arguments,  and with the \textit{general solution} of the system of $3$ \textit{linear}
recursions (\ref{3Recs}) reading then as follows: 
\end{subequations}
\begin{equation}
y_{j}\left( n\right) =\gamma _{j}\left( \overline{u}\right) ^{n}+\alpha
_{j}\left( \widetilde{u}\right) ^{n}\cos \left( nr\pi \right) +\beta
_{j}\left( \widetilde{u}\right) ^{n}\sin \left( nr\pi \right) ~.
\label{3RecsAlfBet}
\end{equation}

This formula features now the $3$ \textit{real }parameters $\overline{u},$ $%
\widetilde{u}>0$ and $r$, related by the formulas (\ref{ajkuur}) to the $9$
coefficients $a_{jk}$ of the system of recursions (\ref{3Recs}); while, of
the $9$ parameters $\gamma _{j}$, $\gamma _{j+}$ and $\gamma _{j-}$, the%
\textit{\ }$3$ parameters $\gamma _{1}$, $\gamma _{1+}$ and $\gamma _{1-}$%
---or, equivalently, the $3$ parameters $\gamma _{1}$, $\alpha _{1}$ and $%
\beta _{1}$ (see (\ref{3RecsAlfBet}))---may be \textit{arbitrarily assigned}%
, while the other $6$ parameters $\gamma _{s}$, $\gamma _{s+}$ and $\gamma
_{s-}$ (with $s=2,3$)---or, equivalently, the $6$ parameters $\gamma _{s}$, $%
\alpha _{s}$ and $\beta _{s}$ (see (\ref{3RecsAlfBet})) with $s=2,3$--- are
determined as the solution of the system of \textit{algebraic} equations (%
\ref{Eqsgam}), implying 
\begin{subequations}
\label{gams+-}
\begin{equation}
\gamma _{s}=\overline{U}_{s}\gamma _{1}~,~~~\gamma _{s\pm }=\widetilde{U}%
_{s\pm }\gamma _{s}~,~~\ s=2,3~,
\end{equation}%
where%
\begin{eqnarray}
\overline{U}_{2} &=&\frac{\left( \overline{u}-a_{33}\right)
a_{21}+a_{23}a_{31}}{\left( \overline{u}-a_{22}\right) \left( \overline{u}%
-a_{33}\right) -a_{23}a_{32}}~, \\
\overline{U}_{3} &=&\frac{\left( \overline{u}-a_{22}\right)
a_{31}+a_{32}a_{21}}{\left( \overline{u}-a_{22}\right) \left( \overline{u}%
-a_{33}\right) -a_{23}a_{31}}~;
\end{eqnarray}%
\begin{eqnarray}
\widetilde{U}_{2\pm } &=&\frac{\left[ \widetilde{u}\exp \left( \pm \mathbf{i}%
\pi r\right) -a_{33}\right] a_{21}+a_{23}a_{31}}{\left[ \widetilde{u}\exp
\left( \pm \mathbf{i}\pi r\right) -a_{22}\right] \left[ \widetilde{u}\exp
\left( \pm \mathbf{i}\pi r\right) -a_{33}\right] -a_{23}a_{32}}~, \\
\widetilde{U}_{3\pm } &=&\frac{\left[ \left( \widetilde{u}\exp \left( \pm 
\mathbf{i}\pi r\right) -a_{22}\right) \right] a_{31}+a_{32}a_{21}}{\left[ 
\widetilde{u}\exp \left( \pm \mathbf{i}\pi r\right) -a_{22}\right] \left[ 
\widetilde{u}\exp \left( \pm \mathbf{i}\pi r\right) -a_{33}\right]
-a_{23}a_{31}}~.
\end{eqnarray}

The last $2$ formulas may be written---at the cost of some trivial if
tedious computations ---in the following more explicit form: 
\begin{eqnarray}
&&\widetilde{U}_{2\pm }=\frac{  \Re  \left[Num\widetilde{U}_{2}\right]\pm \mathbf{i}%
 \Im \left[ Num\widetilde{U}_{2}\right]}{Den\widetilde{U}_{2}}~, \\
&&  \Re \left[  Num\widetilde{U}_{2}\right]=-\left( a_{21}a_{33}-a_{23}a_{31}\right)
\left( a_{22}a_{33}-a_{23}a_{32}\right) -a_{21}\left( a_{22}+a_{33}\right)
\left( \widetilde{u}\right) ^{2}+  \notag \\
&&\left[ \left( 2a_{21}a_{33}-a_{23}a_{31}\right)
a_{22}-a_{23}a_{31}a_{33}+a_{21}\left( -a_{23}a_{32}+\left( a_{33}\right)
^{2}+\left( \widetilde{u}\right) ^{2}\right) \right] \cdot  \notag \\
&&\cdot \widetilde{u}\cos \left( 2\pi r\right) -\left(
a_{21}a_{33}-a_{23}a_{31}\right) \left( \widetilde{u}\right) ^{2}\cos \left(
2\pi r\right) ~, \\
&& \Im Num\widetilde{U}_{2}=\left\{ {}\right. \left[
a_{22}a_{23}a_{31}+a_{23}a_{31}a_{33}-a_{21}\left( -a_{23}a_{32}+\left(
a_{33}\right) ^{2}\right) \right] \widetilde{u}  \notag \\
&&-a_{21}\left( \widetilde{u}\right) ^{3}+2\left(
a_{21}a_{33}-a_{23}a_{31}\right) \widetilde{u}\cos \left( \pi r\right)
\left. {}\right\} \sin \left( 2\pi r\right) ~, \\
&&Den\widetilde{U}_{2}=\left( a_{22}a_{33}-a_{23}a_{32}\right) ^{2}+\left(
a_{22}+a_{33}\right) ^{2}\left( \widetilde{u}\right) ^{2}+\left( \widetilde{u%
}\right) ^{4}+  \notag \\
&&2\left( a_{22}+a_{33}\right) \left( a_{22}a_{33}-a_{23}a_{32}-\left( 
\widetilde{u}\right) ^{2}\right) \widetilde{u}\cos \left( 2\pi r\right) - 
\notag \\
&&2\left( a_{22}a_{33}-a_{23}a_{32}\right) \left( \widetilde{u}\right)
^{2}\cos \left( 2\pi r\right) ~;
\end{eqnarray}%
\begin{eqnarray}
&&\widetilde{U}_{3\pm }=\frac{  \Re \left[  Num\widetilde{U}_{3}\right]\pm \mathbf{i}%
 \Im \left[  Num\widetilde{U}_{3}\right]}{Den\widetilde{U}_{3}}~, \\
&&  \Re \left[ Num\widetilde{U}_{3}\right]=\left( a_{21}a_{32}-a_{31}a_{33}\right)
\left( a_{22}a_{33}-a_{23}a_{32}\right) -\left( a_{22}+a_{33}\right)
a_{31}\left( \widetilde{u}\right) ^{2}+  \notag \\
&&\left\{ -a_{21}\left( a_{22}+a_{33}\right)
a_{32}+2a_{22}a_{31}a_{33}-a_{23}a_{31}a_{32}+a_{31}\left[ \left(
a_{33}\right) ^{2}+a_{31}\left( \widetilde{u}\right) ^{2}\right] \right\}
\cdot  \notag \\
&&\widetilde{\cdot u}\cos \left( \pi r\right) +\left(
a_{21}a_{32}-a_{31}a_{33}\right) \left( \widetilde{u}\right) ^{2}\cos \left(
2\pi r\right) ~, \\
&& \Im \left[ Num\widetilde{U}_{3}\right]=a_{21}\left( a_{22}+a_{33}\right)
a_{32}-a_{23}a_{31}a_{32}-a_{31}\left[ \left( a_{33}\right) ^{2}-\left( 
\widetilde{u}\right) ^{2}\right] -  \notag \\
&&2\left( a_{21}a_{32}-a_{31}a_{33}\right) \widetilde{u}\cos \left( 2\pi
r\right) ~, \\
&&Den\widetilde{U}_{3}=\left( a_{22}a_{33}-a_{23}a_{32}\right) ^{2}+\left(
a_{22}+a_{33}\right) ^{2}\left( \widetilde{u}\right) ^{2}+\left( \widetilde{u%
}\right) ^{4}+  \notag \\
&&2\left( a_{22}+a_{33}\right) \left[ a_{22}a_{33}-a_{23}a_{32}+\left( 
\widetilde{u}\right) ^{2}\right] \widetilde{u}\cos \left( 2\pi r\right) + 
\notag \\
&&2\left( a_{22}a_{33}-a_{23}a_{32}\right) \left( \widetilde{u}\right)
^{2}\cos \left( 2\pi r\right) ~.
\end{eqnarray}%
And obviously these explicit formulas also imply 
\begin{equation}
\gamma _{s}=\overline{U}_{s}\gamma _{1}~,~~~\alpha _{s}=2\left[ \frac{  \Re \left[  Num\widetilde{U}_{2}\right]}{Den\widetilde{U}_{2+}}\right] \gamma _{s}~,~~\beta
_{s}=2\left[ \frac{ \Im \left[ Num\widetilde{U}_{2+}\right]}{Den\widetilde{U}_{2+}}%
\right] \gamma _{s}~,~~s=2,3~.
\end{equation}

\bigskip

\textbf{The initial-values problem}

\bigskip

And finally, in the context of the \textit{initial-values} problem, the $3$
parameters $\gamma _{1}$, $\gamma _{1+}$ and $\gamma _{1-}$ ---or,
equivalently, the $3$ parameters $\gamma _{1}$, $\alpha _{1}$ and $\beta
_{1} $---are determined from the $3$ \textit{initial-values} $y_{j}\left(
0\right) $ as the solutions of the system of $3$ \textit{linear} equations (%
\ref{yjngam}) (or (\ref{3RecsAlfBet})) at $n=0$, clearly implying: $\ $%
\end{subequations}
\begin{subequations}
\begin{equation}
\gamma _{j}+\gamma _{j+}+\gamma _{j-}=y_{j}\left( 0\right) ~,
\end{equation}%
namely%
\begin{eqnarray}
\gamma _{1}+\gamma _{1+}+\gamma _{1-} &=&\gamma _{1}+\alpha _{1}=y_{1}\left(
0\right) ~, \\
\gamma _{2}+\gamma _{2+}+\gamma _{2-} &=&\gamma _{2}+\alpha _{2}=y_{2}\left(
0\right) ~, \\
\gamma _{3}+\gamma _{3+}+\gamma _{3-} &=&\gamma _{3}+\alpha _{3}=y_{3}\left(
0\right) ~,
\end{eqnarray}%
where the $6$ parameters $\gamma _{s}$, $\gamma _{s+}$ and $\gamma _{s-}$
(with $s=2,3$) are given in terms of the $3$ parameters $\gamma _{1}$, $%
\gamma _{1+}$ and $\gamma _{1-}$ by the formulas (\ref{gams+-}) (and
likewise, for the relations of the $9$ parameters $\gamma _{j}$, $\alpha
_{j} $ and $\beta _{j}$ to the initial data $y_{j}\left( 0\right) $, see
below).

These formulas provide both the \textit{general solution} of the \textit{%
linear} system of $3$ recursions (\ref{3Recs}) and of its \textit{%
initial-values} problem (as well, of course, of the marginally more general
system (\ref{3Recsz})). They cannot certainly be considered a new
mathematical finding, being only an exercise in rather elementary algebra.
Yet they display in \textit{explicit} form the---somewhat
nontrivial---phenomenology of the solutions of the system of $3$ recursions (%
\ref{3Recs}); this might be of potential interest in various applicative
sciences for practitioners (who might possibly themselves be less familiar
with the elementary mathematics needed to solve these equations), as well as
in order to identify---again, by familiar techniques well known to
mathematicians (for instance, just changes of variables: see below)---less
trivial systems\ of $3$ (\textit{nonlinear}!) recursions which are also 
\textit{explicitly solvable} and which might therefore be themselves of
applicative interest in various contexts.

\bigskip

\textbf{The qualitative features of these solutions}

\bigskip

Let us outline here the features of these behaviors which are immediate
consequences of the solution formula (\ref{3RecsAlfBet}).

Quite evidently, the evolution of this \textit{general solution} (\ref%
{3RecsAlfBet}) as a function of the \textit{discrete independent variable }$%
n=0,1,2,3...$ depends crucially on the values of the $2$ \textit{real
numbers }$\overline{u}$ and $\widetilde{u}$, and in a more subtle way on the 
\textit{real number} $r$. Indeed, for instance, if both $\left\vert 
\overline{u}\right\vert <1$ and $\left\vert \widetilde{u}\right\vert <1$,
then clearly \textit{all solutions} $y\left( n\right) $ shall steadily
decrease in modulus and eventually \textit{vanish} as $n$ grows: 
\end{subequations}
\begin{equation}
y_{j}\left( n\right) \rightarrow 0~~~\text{as}~~~n\rightarrow \infty
~,~~~j=1,2,3~.
\end{equation}%
And likewise if either \textit{one}, or bo\textit{th}, of the $2$ \textit{%
real numbers }$\overline{u}$ and $\widetilde{u}$ are \textit{larger than }$1$
in \textit{modulus}, then the $3$ solutions $y_{j}\left( n\right) $ shall
generally \textit{all diverge asymptotically (in modulus) }as $n\rightarrow
\infty $.

More interesting is the behavior if $\left\vert \widetilde{u}\right\vert =1$%
\ and either $\overline{u}=0$ or $\overline{u}=1$; this of course happens
provided the $9$ coefficients $a_{jk}$ characterizing the system of $3$
recursions (\ref{3Recs}) satisfy $2$ \textit{restrictions} (see
below)---because then clearly (see (\ref{3RecsAlfBet})) the $3$ components $%
y_{j}\left( n\right) $ of the solution, as $n\rightarrow \infty $, shall
eventually fill the interval from $-\left\vert \alpha _{j}\right\vert
-\left\vert \beta _{j}\right\vert $ to $\left\vert \alpha _{j}\right\vert
+\left\vert \beta _{j}\right\vert $ if $\overline{u}=0$ (or, instead, from $%
\gamma _{j}-\left\vert \alpha _{j}\right\vert -\left\vert \beta
_{j}\right\vert $ to $\gamma _{j}+\left\vert \alpha _{j}\right\vert
+\left\vert \beta _{j}\right\vert $ if $\overline{u}=1$); \textit{unless}
the \textit{real number }$r$ happens to be a \textit{rational number}, 
\begin{subequations}
\begin{equation}
r=2N_{1}/N_{2}~,  \label{rrational}
\end{equation}%
where (above and hereafter) $N_{1}$ and $N_{2}$ are $2$ \textit{arbitrary
nonvanishing integers} (with $N_{2}>2N_{1}>0$, see (\ref{r}) and recall the
restriction (\ref{r})), in which case the \textit{general solution} of the
system of $3$ recursions (\ref{3Recs}) features the remarkable property to
be \textit{isochronous}, \textit{all} its solutions being then \textit{%
periodic} with \textit{period} $N_{2}$, 
\begin{equation}
y_{j}\left( n+N_{2}\right) =y_{j}\left( n\right)
~,~~~j=1,2,3,~~~n=0,1,2,3...~.  \label{Periodic}
\end{equation}%
So this is a property of the system of $3$ recursions (\ref{3Recs}) which
emerges whenever its $9$ coefficients imply the following restrictions on
the $3$ \textit{real numbers }$\overline{u}$, $\widetilde{u}$ and $r$:%
\begin{equation}
\overline{u}=0\text{ or }\overline{u}=1~,~~~\widetilde{u}%
=1~,~~~r=2N_{1}/N_{2}~.
\end{equation}%
It is also clear that this \textit{periodicity} property would also
prevail---possibly up to a \textit{doubling} of the period---if $\overline{u}%
=-1$ (instead of $\overline{u}=1$; while we may ignore the case $\widetilde{u%
}=-1$ since we assumed that $\widetilde{u}>0$).

And our final observation is that the \textit{periodicity properties}
detailed just above get \textit{replaced} by the following properties of 
\textit{asymptotic periodicity,} 
\end{subequations}
\begin{subequations}
\begin{equation}
\underset{n\rightarrow \infty }{Lim}\left[ y_{j}\left( n+N_{2}\right)
-y_{j}\left( n\right) \right] =0~,~~~j=1,2,3,~~~n=0,1,2,3,...~,
\end{equation}%
if---while of course maintaining the requirement $\widetilde{u}=1$--- the 
\textit{restrictions} $\overline{u}=0$ or $\overline{u}=\pm 1$ are replaced\
by the (\textit{less stringent}) requirement%
\begin{equation}
\left\vert \overline{u}\right\vert <1~.
\end{equation}

\bigskip

\textbf{Simple examples}

\bigskip

The key formulas to identify examples are those (see (\ref{ajkuur}))
relating the coefficients $a_{jk}$ to the $3$ parameters $\overline{u},$ $%
\widetilde{u}$ and $r$. For instance if $\overline{u}=0$, $\widetilde{u}=1$
and $r=2N_{1}/N_{2}$, the $3$ \textit{constraints} on the $9$ coefficients $%
a_{jk}$ which are \textit{necessary and sufficient} to imply that the
corresponding system of $3$ \textit{linear homogeneous recurrences} (\ref%
{3Recs}) be \textit{isochronous} with period $N_{2}$ clearly read as
follows: 
\end{subequations}
\begin{subequations}
\label{aa}
\begin{eqnarray}
&&a_{11}+a_{22}+a_{33}=2\cos \left( 2\pi N_{1}/N_{2}\right) ~,  \label{aa1}
\\
&&a_{11}a_{22}+a_{22}a_{33}+a_{33}a_{11}-a_{12}a_{21}-a_{13}a_{31}-a_{23}a_{32}=1~,
\label{aa2} \\
&&a_{11}a_{22}a_{33}+a_{12}a_{23}a_{31}+a_{13}a_{21}a_{32}-  \notag \\
&&a_{13}~a_{31}a_{22}-a_{12}a_{21}a_{33}-a_{23}a_{32}a_{11}=0~.  \label{aa3}
\end{eqnarray}%
This implies for instance the somewhat remarkable result---corresponding to
the assignment $N_{1}/N_{2}=1/2$---that any system of $3$ \textit{linear
homogeneous recurrences} (\ref{3Recs}) such that its $9$ coefficients $%
a_{jk} $ satisfy the following $3$ \textit{constraints}, 
\end{subequations}
\begin{subequations}
\label{aaa}
\begin{eqnarray}
&&a_{11}+a_{22}+a_{33}=-2~,  \label{aaa1} \\
&&a_{11}a_{22}+a_{22}a_{33}+a_{33}a_{11}-a_{12}a_{21}-a_{13}a_{31}-a_{23}a_{32}=1~,
\label{aaa2} \\
&&a_{11}a_{22}a_{33}+a_{12}a_{23}a_{31}+a_{13}a_{21}a_{32}-  \notag \\
&&a_{13}~a_{31}a_{22}-a_{12}a_{21}a_{33}-a_{23}a_{32}a_{11}=0~,  \label{aaa3}
\end{eqnarray}%
shall be \textit{isochronous} with period $2$: 
\end{subequations}
\begin{equation}
y_{j}\left( n+2\right) =y_{j}\left( n\right) ~,~~~j=1,2,3,~~~n=0,1,2,3,...~.
\label{Period2}
\end{equation}%
Note that the $3$ eqs. (\ref{aaa}) may be \textit{explicitly} solved, for
instance for $a_{11}$, $a_{12}$ and $a_{13}$: 
\begin{subequations}
\label{Aaaa}
\begin{eqnarray}
&&a_{11}=-a_{22}-a_{33}-2,  \label{a11} \\
&&a_{12}=\left[ {}\right. a_{21}a_{32}-a_{22}a_{31}+\left(
a_{21}a_{22}a_{32}-2a_{23}a_{31}a_{32}-\left( a_{22}\right)
^{2}a_{31}\right) a_{22}  \notag \\
&&+(a_{21}a_{23}+a_{21}a_{22}-a_{23}a_{31}+a_{21}a_{33})a_{32}a_{33}\left.
{} \right] /A_{123}~,  \label{a12} \\
&&a_{13}=\left[ {}\right.
a_{21}a_{33}-a_{23}a_{31}+a_{21}a_{22}a_{23}a_{32}-\left( a_{22}\right)
^{2}a_{23}a_{31}-\left( a_{23}\right) ^{2}a_{31}a_{32}-  \notag \\
&&\left( a_{21}\left( a_{33}\right)
^{2}+2a_{21}a_{23}a_{32}+a_{22}a_{23}a_{31}-a_{23}a_{31}a_{33}\right)
a_{33}\left. {}\right] /A_{123}~,  \label{a13} \\
&&A_{123}=-a_{21}\left( a_{21}a_{32}+a_{31}a_{33}\right) +\left(
a_{21}a_{22}+a_{23}a_{31}\right) a_{31}~.  \label{A123}
\end{eqnarray}%
We are of course assuming that $A_{123}$ does \textit{not} vanish.

Moreover, the same \textit{isochrony }outcome (\ref{Period2}) shall emerge
in the \textit{alternative} case when the $9$ coefficients $a_{jk}$ satisfy
the following $3$ \textit{alternative constraints}: 
\end{subequations}
\begin{subequations}
\begin{eqnarray}
&&a_{11}+a_{22}+a_{33}=1~, \\
&&a_{11}a_{22}+a_{22}a_{33}+a_{33}a_{11}-a_{12}a_{21}-a_{13}a_{31}-a_{23}a_{32}=1~,
\\
&&a_{11}a_{22}a_{33}+a_{12}a_{23}a_{31}+a_{13}a_{21}a_{32}-  \notag \\
&&a_{13}a_{31}a_{22}-a_{12}a_{21}a_{33}-a_{23}a_{32}a_{11}=1~,
\end{eqnarray}%
implying $\overline{u}=\widetilde{u}=1$ and $r=N_{1}/N_{2}=1/2$ hence $\cos
\left( \pi r\right) =0$; and these equations may be again solved for $a_{11}$%
, $a_{12}$ and $a_{13}$: 
\begin{eqnarray}
&&a_{11}=1-a_{22}-a_{33}~, \\
&&a_{12}=\left\{ {}\right. a_{21}+a_{21}a_{32}-a_{22}a_{21}+  \notag \\
&&\left( a_{22}\right) ^{2}a_{21}-a_{21}a_{22}a_{32}+a_{23}a_{21}a_{32}- 
\notag \\
&&2a_{22}a_{23}a_{21}a_{32}+a_{21}\left[ \left( a_{22}\right)
^{2}a_{32}+a_{23}\left( a_{32}\right) ^{2}+a_{32}\left( a_{33}\right) ^{2}%
\right] +  \notag \\
&&a_{21}a_{33}\left( a_{22}-a_{32}\right) -a_{23}a_{21}a_{32}a_{33}\left.
{}\right\} /A_{123}~, \\
&&a_{13}=-\left[ {}\right. a_{21}+a_{23}a_{21}-a_{22}a_{23}a_{21}+\left(
a_{22}\right) ^{2}a_{23}a_{21}+  \notag \\
&&a_{21}a_{23}a_{32}-a_{21}a_{22}a_{23}a_{32}+\left( a_{23}\right)
^{2}a_{21}a_{32}-a_{21}a_{33}-  \notag \\
&&a_{23}a_{21}a_{33}+a_{22}a_{23}a_{21}a_{33}-2a_{21}a_{23}a_{32}a_{33}+ 
\notag \\
&&a_{21}\left( a_{33}\right) ^{2}+a_{23}a_{21}\left( a_{33}\right)
^{3}-a_{21}\left( a_{33}\right) ^{3}\left. {}\right] /A_{123}~.
\end{eqnarray}%
And there are of course other simple cases which the interested reader may
well enjoy to identify and possibly use.

So much for the simple case of a system of $3$ \textit{first-order linear}
recurrences. Let us now use these findings to explore the much larger
universe of systems of $3$ \textit{first-order nonlinear} recurrences.

A step in this direction is to introduce a simple change of dependent
variables.

\bigskip

\textbf{A change of dependent variables}

\bigskip

So we now introduce a \textit{new} set of $3$ dependent variables, related
to the dependent variables $y_{j}\left( n\right) $ used above by the
following formulas: 
\end{subequations}
\begin{equation}
x_{j}\left( n\right) =\frac{b_{j0}+b_{j1}y_{1}\left( n\right)
+b_{j2}y_{2}\left( n\right) +b_{j3}y_{3}\left( n\right) }{%
c_{j0}+c_{j1}y_{1}\left( n\right) +c_{j2}y_{2}\left( n\right)
+c_{j3}y_{3}\left( n\right) }~,~~~j=1,2,3~.  \label{xy}
\end{equation}

These formulas introduce $2\cdot 3\cdot 4=24$ \textit{a priori} free \textit{%
real} parameters $b_{j\ell }$ and $c_{j\ell }$. Once their values are set,
the $n$-evolution of the $3$ new dependent variables $x_{j}\left( n\right) $
is of course \textit{explicitly} determined by the $n$-evolution of the $3$
dependent variables $y_{j}\left( n\right) $, as discussed above.

These formulas can be explicitly \textit{inverted}. To do so it is
convenient to introduce the following matrix/vector notation: 
\begin{subequations}
\begin{equation}
\mathbf{x}\left( n\right) =\left( 
\begin{array}{c}
x_{1}\left( n\right) \\ 
x_{2}\left( n\right) \\ 
x_{3}\left( n\right)%
\end{array}%
\right) ~,~~~\mathbf{y}\left( n\right) =\left( 
\begin{array}{c}
y_{1}\left( n\right) \\ 
y_{2}\left( n\right) \\ 
y_{3}\left( n\right)%
\end{array}%
\right) ~,  \label{Vecxy}
\end{equation}%
\begin{equation}
\mathbf{b}=\left( 
\begin{array}{c}
b_{10} \\ 
b_{20} \\ 
b_{30}%
\end{array}%
\right) ~,~~~\mathbf{c}=\left( 
\begin{array}{c}
c_{10} \\ 
c_{20} \\ 
c_{30}%
\end{array}%
\right) ~,  \label{Vecbc}
\end{equation}%
\begin{equation}
\mathbf{B=}\left( 
\begin{array}{ccc}
b_{11} & b_{12} & b_{13} \\ 
b_{21} & b_{22} & b_{23} \\ 
b_{31} & b_{32} & b_{33}%
\end{array}%
\right) ~,~~~\mathbf{C=}\left( 
\begin{array}{ccc}
c_{11} & c_{12} & c_{13} \\ 
c_{21} & c_{22} & c_{23} \\ 
c_{31} & c_{32} & c_{33}%
\end{array}%
\right) ~.  \label{MatBC}
\end{equation}%
Hence 
\end{subequations}
\begin{eqnarray}
&&\left( 
\begin{array}{ccc}
c_{11}x_{1}\left( n\right) -b_{11} & c_{12}x_{1}\left( n\right) -b_{12} & 
c_{13}x_{1}\left( n\right) -b_{13} \\ 
c_{21}x_{2}\left( n\right) -b_{21} & c_{22}x_{2}\left( n\right) -b_{21} & 
c_{23}x_{2}\left( n\right) -b_{23} \\ 
c_{31}x_{3}\left( n\right) -b_{31} & c_{32}x_{3}\left( n\right) -b_{31} & 
c_{33}x_{3}\left( n\right) -b_{33}%
\end{array}%
\right) \left( 
\begin{array}{c}
y_{1}\left( n\right) \\ 
y_{2}\left( n\right) \\ 
y_{3}\left( n\right)%
\end{array}%
\right)  \notag \\
&=&\left( 
\begin{array}{c}
b_{10}-c_{10}x_{1}\left( n\right) \\ 
b_{20}-c_{20}x_{2}\left( n\right) \\ 
b_{30}-c_{30}x_{3}\left( n\right)%
\end{array}%
\right)
\end{eqnarray}%
implying 
\begin{subequations}
\label{ynxn}
\begin{eqnarray}
\left( 
\begin{array}{c}
y_{1}\left( n\right) \\ 
y_{2}\left( n\right) \\ 
y_{3}\left( n\right)%
\end{array}%
\right) &=&\left( 
\begin{array}{ccc}
c_{11}x_{1}\left( n\right) -b_{11} & c_{12}x_{1}\left( n\right) -b_{12} & 
c_{13}x_{1}\left( n\right) -b_{13} \\ 
c_{21}x_{2}\left( n\right) -b_{21} & c_{22}x_{2}\left( n\right) -b_{21} & 
c_{23}x_{2}\left( n\right) -b_{23} \\ 
c_{31}x_{3}\left( n\right) -b_{31} & c_{32}x_{3}\left( n\right) -b_{31} & 
c_{33}x_{3}\left( n\right) -b_{33}%
\end{array}%
\right) ^{-1}\cdot  \notag \\
&&\left( 
\begin{array}{c}
b_{10}-c_{10}x_{1}\left( n\right) \\ 
b_{20}-c_{20}x_{2}\left( n\right) \\ 
b_{30}-c_{30}x_{3}\left( n\right)%
\end{array}%
\right)  \label{yx}
\end{eqnarray}%
namely%
\begin{equation}
y_{j}\left( n\right) =\frac{P_{4}^{\left( j\right) }\left( x_{1}\left(
n\right) ,x_{2}\left( n\right) ,x_{3}\left( n\right) \right) }{P_{3}\left(
x_{1}\left( n\right) ,x_{2}\left( n\right) ,x_{3}\left( n\right) \right) }~,
\label{yjn}
\end{equation}%
where the notation $P_{4}^{\left( j\right) }\left( x_{1},x_{2},x_{3}\right) $
respectively $P_{3}\left( x_{1},x_{2},x_{3}\right) $ denote the $3$
polynomials of $4$-th degree respectively the single polynomial of $3$-rd
degree in the $3$ variables $x_{1},x_{2},x_{3}$ appearing in the right-hand
sides of the formula (\ref{ynxn}).

\bigskip

\textbf{A new \textit{solvable} system of }$3$\textbf{\ first-order \textit{%
nonlinear} recurrences}

\bigskip

The formulas (\ref{xy}) also imply 
\end{subequations}
\begin{subequations}
\begin{equation}
x_{j}\left( n+1\right) =\frac{b_{j0}+b_{j1}y_{1}\left( n+1\right)
+b_{j2}y_{2}\left( n+1\right) +b_{j3}y_{3}\left( n+1\right) }{%
c_{j0}+c_{j1}y_{1}\left( n+1\right) +c_{j2}y_{2}\left( n+1\right)
+c_{j3}y_{3}\left( n+1\right) }~,
\end{equation}%
hence, via (\ref{3Recs}),%
\begin{equation}
x_{j}\left( n+1\right) =\frac{b_{j0}+f_{j1}y_{1}\left( n\right)
+f_{j2}y_{2}\left( n\right) +f_{j3}y_{3}\left( n\right) }{%
c_{j0}+g_{j1}y_{1}\left( n\right) +g_{j2}y_{2}\left( n\right)
+g_{j3}y_{3}\left( n\right) }~,
\end{equation}%
where%
\begin{equation}
f_{jk}=\sum_{s=1}^{3}\left[ a_{js}b_{sk}\right] ~,~~~g_{jk}=\sum_{s=1}^{3}%
\left[ a_{js}c_{sk}\right] ~;
\end{equation}%
hence (via (\ref{yjn})) 
\end{subequations}
\begin{subequations}
\label{NonLin3Recs}
\begin{equation}
x_{j}\left( n+1\right) =\frac{F_{4}^{\left( j\right) }\left( x_{1}\left(
n\right) ,x_{2}\left( n\right) ,x_{3}\left( n\right) \right) }{G_{4}^{\left(
j\right) }\left( x_{1}\left( n\right) ,x_{2}\left( n\right) ,x_{3}\left(
n\right) \right) }~,~~~j=1,2,3~,  \label{xjneq}
\end{equation}%
where 
\begin{equation}
F_{4}^{\left( j\right) }\left( x_{1},x_{2},x_{3}\right) =b_{j0}P_{3}\left(
x_{1},x_{2},x_{3}\right) +\sum_{k=1}^{3}f_{jk}P_{4}^{\left( k\right) }\left(
x_{1},x_{2},x_{3}\right) ~,  \label{F4j}
\end{equation}%
\begin{equation}
G_{4}^{\left( j\right) }\left( x_{1},x_{2},x_{3}\right) =c_{j0}P_{3}\left(
x_{1},x_{2},x_{3}\right) +\sum_{k=1}^{3}g_{jk}P_{4}^{\left( k\right) }\left(
x_{1},x_{2},x_{3}\right) ~.  \label{G4j}
\end{equation}

This new system of $3$ \textit{nonlinear }first-order recursions (\ref%
{NonLin3Recs}) is \textit{explicitly solvable} provided the $3\cdot 2=6$
polynomials of $4$-th degree $F_{4}^{\left( j\right) }\left(
x_{1},x_{2},x_{3}\right) $ and $G_{4}^{\left( j\right) }\left(
x_{1},x_{2},x_{3}\right) $ featured by its right-hand sides are defined as
detailed by the above developments, which entail that their $2\cdot 3\cdot
32=192$ coefficients are given by \textit{explicitly} known, if rather
complicated, sequences of formulas in terms of the $9+24=33$ \textit{a priori%
} free parameters $a_{jk},\ b_{j\ell }$ and $c_{j\ell }$ which characterized
the change of variables (\ref{xy}), as well as the system of $3$ \textit{%
linear }first-order (\ref{3Recs}) satisfied by the dependent variables $%
y_{j}\left( n\right) $. This of course also implies that, if the $9$
coefficients $a_{jk}$ satisfy moreover the appropriate \textit{few}
restrictions discussed above, which are \textit{sufficient} to ensure that
the behavior of the solutions $y_{j}\left( n\right) $ be remarkably
simple---for instance, \textit{isochronous }or \textit{aymptotically
isochronous}---then the \textit{same} behavior shall as well characterize
the solutions of the more complicated system of $3$ first-order \textit{%
nonlinear} recursions (\ref{3Recs}).

There are of course \textit{subcases} when the unbalance among the number of
freely assignable parameters $a_{jk},\ b_{j\ell }$ and $c_{j\ell }$ and the
number of coefficients featured by the recursions (\ref{3Recs}) may
decrease: for instance, the subcase when (in the change of variables (\ref%
{xy})) the $9$ parameters $b_{jk}$ \textit{all} vanish, $b_{jk}=0$, the new
system of recursions satisfied by the dependent variables $x_{j}\left(
n\right) $ reads clearly as follows: 
\end{subequations}
\begin{equation}
x_{j}\left( n+1\right) =\frac{F_{3}^{\left( j\right) }\left( x_{1}\left(
n\right) ,x_{2}\left( n\right) ,x_{3}\left( n\right) \right) }{G_{4}^{\left(
j\right) }\left( x_{1}\left( n\right) ,x_{2}\left( n\right) ,x_{3}\left(
n\right) \right) }~,~~~j=1,2,3~,  \label{3NewRecs}
\end{equation}%
featuring the $3$ polynomials of $3$-rd degree $F_{3}^{\left( j\right)
}\left( x_{1},x_{2},x_{3}\right) $ (rather than $4$-th degree polynomials of 
$4$-rd degree $F_{4}^{\left( j\right) }\left( x_{1},x_{2},x_{3}\right) $) in
the numerators of their right-hand sides, hence altogether $3\cdot
(17+32)=147$ coefficients (instead of $192$); while the number of \textit{a
priori} freely disposable parameters $a_{jk},\ b_{j\ell }$ and $c_{j\ell }~$%
is reduced from $9+24=33$ to $9+3+12=24$. And of course other reductions may
be envisaged. We leave their exploration to interested researchers; as well
as extensions, see below.

\bigskip

\textbf{Conclusions and outlook}

\bigskip

In the right-hand sides of the recursions (\ref{3NewRecs}), each of the $3$
polynomials $P_{4}^{\left( j\right) }\left( x_{1},x_{2},x_{3}\right) $ of $4$%
-th degree features of course $32$ coefficients, while the polynomial of $3$%
-rd degree $P_{3}\left( x_{1},x_{2},x_{3}\right) $ features $17$\
coefficients; it would make hardly any sense to try and report here the
corresponding $32\cdot 3+17=113$ formulas expressing these coefficients in
terms of the $24$ \textit{a priori} free \textit{real} parameters $b_{j\ell
} $ and $c_{j\ell }$; but note that obviously this implies that these $113$
coefficients are severely constrained. And clearly any attempt to identify 
\textit{explicitly} these \textit{constraints} looks doomed to failure.
While it is of course obvious that \textit{not all} these coefficients play
a \textit{significant} role; it is for instance plain that any one of these
coefficients---unless it vanished to begin with---might be replaced---in
each of (the right-hand sides of) the $3$ recursions (\ref{3NewRecs}) by the
number $1$, by simply dividing both the numerator and the denominator
featured by the right-hand side each of the $3$ eqs. (\ref{3NewRecs}) by
that number (an operation that leaves of course invariant those equations).

And clearly the situation would be even less manageable in the more general
case (\ref{NonLin3Recs}) featuring $192$ \textit{a priori} free coefficients.

On the other hand it would be a relatively easy---if possibly somewhat
cumbersome---\textit{numerical} task to \textit{compute}---using the
formulas displayed above---the values of all the $113$ parameters featured
by the right-hand sides of the system of $3$ recursions (\ref{3NewRecs})
which correspond to any chosen set of the $9+24=33$ \textit{a priori} free
parameters $a_{jk},$\ $b_{j\ell }$ and $c_{j\ell }$, and to thereby identify
a subclass of systems of $3$ \textit{nonlinear} recursions (\ref{3NewRecs})
which feature a \textit{predictable} evolution; and likewise for the system (%
\ref{NonLin3Recs}) featuring $192$ \textit{a priori} free coefficients.
These findings are therefore likely\ to be quite useful in some \textit{%
applicative} contexts. Indeed let us recall that recursions such as (\ref%
{NonLin3Recs}) or (\ref{3NewRecs}) may have $2$ quite different kinds of 
\textit{applicative} relevance: when they provide a mathematical description
of some phenomenon---be it natural, economic, epidemiological,...--- whose
evolution in the ticking time $n$ one is interested to understand; but also
when one is rather interested to \textit{identify} such a definite \textit{%
context}---or possibly to \textit{manufacture} some specific\textit{\
mechanism }or \textit{instrument}---meant to produce a \textit{desired
outcome}: for instance, a \textit{periodic} behavior of the $3$ dependent
variables $x_{j}\left( n\right) $ as functions of the ticking time $n$, or
instead a behavior leading to their \textit{asymptotic vanishing} as $%
n\rightarrow \infty $. The formulas reported above clearly provide an 
\textit{explicit}---if possibly a bit computationally cumbersome---avenue to 
\textit{manage} this second kind of tasks.

It is my intention to report these findings, as well as the somewhat
analogous ones---see \cite{1} \cite{2}---at the meeting which will take
place in a few days at the University of Rome 1 "La Sapienza" to celebrate
the $70$-th birth-date of Paolo\ Maria Santini: a lifelong friend, who got
his \textit{laurea} in physics at this University writing a thesis under my
supervision, then became eventually a colleague at this University, and is
now joining me as a State pensioner (almost all Universities in Italy are
State Universities).

The simple-minded approach described in this paper may obviously be extended
in various directions---for instance, \textit{more dependent variables} than 
$3$, or \textit{more independent variables} than just $1$; but this will be
tasks for others, as I plan to devote the last period of my life---be they
years, or days---to try and produce a terse autobiography, to be hopefully
entitled \textit{My long life as a scientist and a "pacifist"}.

\end{document}